\documentclass[12pt,reqno]{amsart}
\usepackage[round]{natbib}
\usepackage{amsfonts}
\usepackage{pdfsync}
\usepackage{amssymb, amsmath}
\usepackage{graphicx,graphics,xcolor,epsfig,setspace}
\usepackage[section]{placeins}

\def\sqr#1#2{{\vcenter{\vbox{\hrule height.#2pt
        \hbox{\vrule width.#2pt height#1pt \kern#2pt
        \vrule width.#2pt}
        \hrule height.#2pt}}}}

\newcommand{\nc}{\newcommand}
\nc{\parent}[1]{$[\![#1]\!]$}
\newtheorem{theorem}{Theorem}[section]
\newtheorem{lemma}{Lemma}[section]

\newtheorem{remark}{Remark}
\newtheorem{definition}{Definition}[section]
\newtheorem{assumption}{Assumption}[section]

\nc{\cadlag}{c\`{a}dl\`{a}g } \nc{\ba}{\begin{array}}
\nc{\ea}{\end{array}} \nc{\be}{\begin{equation}}
\nc{\ee}{\end{equation}} \nc{\bea}{\begin{eqnarray}}
\nc{\eea}{\end{eqnarray}} \nc{\bean}{\begin{eqnarray*}}
\nc{\eean}{\end{eqnarray*}} \nc{\bu}{\bullet} \nc{\nn}{\nonumber}
\nc{\cA}{{\mathcal A}} \nc{\cB}{{\mathcal B}} \nc{\cC}{{\mathcal C}}\nc{\cX}{{\mathcal X}}
\nc{\cD}{{\mathcal D}} \nc{\bbD}{\mathbb{D}}\nc{\bbH}{\mathbb{H}}
\nc{\bbF}{\mathbb{F}}\nc{\bbG}{\mathbb{G}}\nc{\cG}{{\mathcal G}} \nc{\cF}{{\mathcal F}}
\nc{\cS}{{\mathcal S}} \nc{\cU}{{\mathcal U}} \nc{\cH}{{\mathcal H}}
\nc{\cK}{{\mathcal K}} \nc{\cL}{{\mathcal L}} \nc{\cM}{{\mathcal M}}
\nc{\cO}{{\mathcal O}} \nc{\cP}{{\mathcal P}} \nc{\cQ}{{\mathcal Q}}\nc{\bbE}{\mathbb{E}}
\nc{\bbN}{\mathbb{N}}
\nc{\bbEQ}{\mathbb{E}_{\mathbb{Q}}} \nc{\eps}{\varepsilon}
\nc{\bbEP}{\mathbb{E}_{\mathbb{P}}}\nc{\bbL}{\mathbb{L}}
\nc{\bbP}{\mathbb{P}} \nc{\bbQ}{\mathbb{Q}} \nc{\del}{\partial}
\nc{\Om}{\Omega} \nc{\om}{\omega} \nc{\bbR}{\mathbb{R}}
\nc{\bbC}{\mathbb{C}} \nc{\bfr}{\begin{flushright}}
\nc{\efr}{\end{flushright}} \nc{\dXt}{\Delta X_{t}} \nc{\dXs}{\Delta
X_{s}} \nc{\bs}{\blacksquare} \nc{\dX}{\Delta X} \nc{\dY}{\Delta Y}
\nc{\dnkx}{\left(X(T^{n}_{k})-X(T^{n}_{k-1})\right)}
\nc{\esssup}{\mathrm{ess}\mbox{ }\mathrm{sup}}
\nc{\essinf}{\mathrm{ess}\mbox{ } \mathrm{inf}}
\nc{\dhats}{\widehat{\delta_s}} \nc{\tX}{\tilde{X}}
\nc{\tZ}{\tilde{Z}}
\nc{\what}{\widehat}
 \nc{\half}{\frac{1}{2}}

\def\rar{\rightarrow}

\nc{\chf}{\mbox{$\mathbf1$}} \nc{\eid}{\stackrel{d}{=}}

\begin{document}
\title{Risk aversion of insider and dynamic asymmetric information.}
\author{Albina Danilova}
\address{Department of Mathematics, London School of Economics and Political Science, 10 Houghton St., London, WC2A 2AE}
\email{a.danilova@lse.ac.uk}
\author{Valentin Lizhdvoy}
\address{Department of Mathematics, Higher School of Economics}
\email{valentin.lizhdvoj@mail.ru}
\date{\today}
\begin{abstract}

This paper studies a Kyle-Back model with a risk-averse insider possessing exponential utility and a dynamic stochastic signal about the asset's terminal fundamental value. While the existing literature considers either risk-neutral insiders with dynamic signals or risk-averse insiders with static signals, we establish equilibrium when both features are present. Our approach imposes no restrictions on the magnitude of the risk aversion parameter, extending beyond previous work that requires sufficiently small risk aversion. We employ a weak conditioning methodology to construct a Schr\"{o}dinger bridge between the insider's signal and the asset price process, an approach that naturally accommodates stochastic signal evolution and removes risk aversion constraints.

We derive necessary conditions for equilibrium, showing that the optimal insider strategy must be continuous with bounded variation. Under these conditions, we characterize the market-maker pricing rule and insider strategy that achieve equilibrium. We obtain explicit closed-form solutions for important cases including deterministic and quadratic signal volatilities, demonstrating the tractability of our framework.
    
\end{abstract}
\maketitle

\newpage

\section{Introduction}
The canonical model of markets with asymmetric information is due to \cite{Kyle}, which introduced a discrete-time model of insider trading and derived its continuous-time equilibrium as a limiting case. The continuous-time framework was formalized by \cite{Back92}, establishing the Kyle-Back model. In this type of models there are typically three types of agents participating in the market: non-strategic noise traders, a strategic informed trader (insider) with private information regarding the future fundamental value of the asset, and risk-neutral market makers competing for the total demand. The goal of market makers is to set the pricing rule so that the resulting price is rational. The objective of the insider is to maximize her expected utility given the pricing rule set by the market makers. Thus, this type of modeling can be viewed as a game with asymmetric information between the market makers and the insider and the goal is to find an equilibrium of this game.

The majority of papers in the Kyle-Back literature consider the case when the insider is risk-neutral (see, e.g., \cite{HSmult}, \cite{BCWmult},
\cite{CD-GKB}). By contrast, the setting of a risk-averse insider with non-linear utility has received significantly less attention despite its economic importance. Early work on risk-averse insiders focused exclusively on models with Gaussian signals (\cite{Bar02}, \cite{ChoRA}), a significantly restrictive assumption. This limitation was relaxed in the work of \cite{BERA23}, who used a fixed point approach coupled with the Fokker-Planck equation and a quasilinear PDE to study models with non-Gaussian insider signals and proved the existence of equilibrium for a broad class of such signals. However, only static type of insider signal was discussed in their work, where the insider knows the asset fundamental value from the onset of trading. Moreover, their approach requires the risk aversion parameter to be sufficiently small for the contraction mapping argument to succeed.  \cite{BERA24} extended this framework to a multidimensional setting with multiple risky assets, again using the fixed-point methodology with static signals and restricted risk aversion.

The cases of dynamic insider signal, when the fundamental value of the traded asset is revealed through a stochastic process, have so far been studied only for the risk-neutral insider (\cite{BP98}, \cite{D}, \cite{CCDbp}, \cite{CCDdef}).

To the best of our knowledge, this paper is the first analysis of a Kyle-Back type model with risk-averse insider having exponential utility function and stochastic dynamic signal about the asset fundamental value. Thus, such framework generalizes the previous results for the static insider signal making them a particular case of our setting. Moreover, our method imposes no restrictions on the magnitude of the risk aversion parameter $\gamma$, whereas the fixed-point approach requires it to be sufficiently small.

A key contribution of our work lies in the methodological approach and in the technique used to prove the existence of the equilibrium. Whereas \cite{BERA23} and \cite{BERA24} relied on a fixed point construction coupling the Fokker-Planck equation with a quasilinear PDE, subject to an optimal transport constraint at maturity, our proof is based on the weak conditioning approach. This method constructs a Schr\"{o}dinger bridge via weak conditioning between the stochastic processes representing insider signal and the asset price process, closely related to entropic optimal transport on path space. This approach naturally accommodates the evolution of information over time, while extending the fixed-point methodology to handle stochastic signal dynamics would require additional technical machinery.

Our main results can be summarized as follows. First, we characterize the market-maker weight function and insider trading strategy that yield the model equilibrium. The existence of equilibrium is demonstrated by constructing a Schrodinger bridge via weak conditioning between the stochastic processes representing the insider signal and the asset price process. Second, we show that the optimal insider strategy must be continuous with bounded variation and that in equilibrium there should be no jump in the asset price and at maturity the price process converges to the signal process. Third, we obtain explicit closed-form equilibrium solutions in particular cases including deterministic and quadratic signal volatilities, enabling comparative statics analysis across different levels of risk aversion.

The paper is structured in the following way. Section 2 introduces the model, specifies the market participants and formulates the market-maker pricing rule along with the insider's utility optimization problem. Section 3 defines admissible pricing rules, admissible insider strategies and model equilibrium. Section 4 presents the main theoretical results of the work -- the Theorem \ref{th:main} and Lemma \ref{th:opt}, which identify the necessary conditions for the equilibrium and describe under what pricing rule and insider strategy it can be achieved. Section 5 illustrates the theoretical results from Section 4 with particular cases, where an equilibrium can be obtained in closed form.

\section{Description of the market model} \label{Market Model}

{In our model we will assume that all processes are defined on the filtered probability space
$$\left(\Omega, \cF, \{\cF_t\}_{t\in[0,1]}, \bbQ^{base}\right)$$

satisfying the usual conditions. Moreover, this space is assumed to be large enough to support two independent Brownian motions $B$  and $\beta$ as well as a standard normal random variable, $Z_0$, independent of $\sigma(B,\beta)$. 

The financial market we study consists of a risk-less and a risky asset, both traded continuously on the interval $[0,1]$. The price of the  risk-less asset is normalized to be a constant, whereas the price of the risky asset is determined in the equilibrium. At the time $1$ the fundamental value of the risky asset, $V$, will be released.  It is given by  $V  = Z_1$ where 
\be \label{e:SDE_Z}
    Z_t =\eta_{V(t)}
    \ee
    with $\eta$ being the unique strong solution of
   \be \label{e:SDE_eta}
    \eta_t = \int_0^t a(s, \eta_s)d\beta_s. 
    \ee 
  In the rest of the paper we impose the following conditions on the structure of the fundamental signal:
  
\begin{assumption}\label{a:Z}
There exists unique strong solution to (\ref{e:SDE_eta}) on $[0,1]$ with the  state space $I:=(l,u)$ with $l,u \in \bar{\bbR}$ and $0\in I$ that admits a transition density and 
    \begin{enumerate}
    \item $a \in C^{1, 2}([0,1],I)$ is positive,
   satisfies 
    \be\label{e:PDEa}
     \frac{a_t}{a^2}(t,x)+\frac{a_{xx}}{2}(t,x)=-\gamma
    \ee
    and  for any $t$ the function $\int_0^x\frac{1}{a(t, y)}dy, \hspace{2mm} x \in I$ has the whole $\mathbb{R}$ as its range.

     \item $\sigma(s)$ is continuous on $[0, 1]$ and is separated from 0. Moreover, it satisfies the following conditions:
      \begin{enumerate}
      \item $V(t): = v_0+\int_0^t \sigma^2(s)ds, v_0 \geq 0$ satisfies $V(1) = 1$ and $V(t)>t$ on $[0,1)$,
    \item $\lim_{t \to 1} D^2(t)\Lambda(t)\log{\Lambda(t)} = 0$ where $D(t) = \exp\left\{-\int_0^t \frac{1}{V(s)-s}ds\right\}$ and $\Lambda(t) = \int_0^t \frac{1+\sigma^2(s)}{D^2(s)}ds$.    
    \end{enumerate}
    \end{enumerate}
\end{assumption}
\begin{remark}\label{Signal} Note that the condition (\ref{e:PDEa}) does not necessarily require that the signal must have the volatility satisfying the PDE. Indeed, as it is shown in Section \ref{ch:DVS} any signal with deterministic volatility can be represented as a signal satisfying condition (\ref{e:PDEa}). Thus, this constraint postulates that the signal can be represented in this form rather than requiring that it is given in this form.
\end{remark}

\begin{remark}\label{TimeCh}
Note that the assumption that $\eta_0 = 0$ is without loss of generality. Indeed, if $\eta_0$ is some constant different from $0$, one can shift the process $\eta_t$ by the $\eta_0$ and modify $a(t, x)$ to  $a(t, x-\eta_0)$.  Then the process $Z$ will have the representation given by  (\ref{e:SDE_Z})-(\ref{e:SDE_eta}).

Moreover, as
$$\eta_{V(t)} = \int_0^{V(t)}a(s, \eta_{s})d\beta_{s},$$
 by employing a time change one can obtain
$$dZ_s = \sigma(s)a(V(s), Z_s)d\tilde{\beta}_s, $$
where $\tilde{\beta}_s$ is the time-changed Brownian motion defined by $\tilde{\beta}_s = \int_0^s\frac{1}{\sigma(u)}d\beta_{V(u)}$.

This yields
$$ Z_t = Z_0 + \int_0^t \sigma(s)a(V(s), Z_s)d\tilde{\beta}_s,
\hspace{5mm} Z_0 = \eta_{V(0)} = \int_0^{V(0)} a(t, \eta_t)d\beta_t,$$
which can be viewed as an alternative representation of the insider signal.
\end{remark}

There are three types of agents populating the market: noise trader, market maker and insider. They are differentiated not only by their optimization problems, but also by the information they have access to, hence by filtrations their actions are adapted to. In particular:

\textbf{\emph{Noise  Traders}} trade for reasons other than maximizing their utilities, and we assume that their cumulative demand follows a standard Brownian motion, $B$.

\textbf{\emph{Market Maker}} observes total cumulated orders,  $Y_t = \theta_t + B_t$, where $ \theta_t$ is the insider's cumulated order by time $t$.
The market maker's filtration at time $t$, $\cF_t^{M}$, is defined as $\cF_t^{M}:=\cF_t^Y$ for $t\in[0,1)$ and $\cF_1^{M}:=\cF_1^{Y,Z}$. 

The market maker sets the asset price, $P$, which in principle can depend on the whole path of $Y$. We will restrict  our attention to the price processes of the form  
$$
P_t = \xi_t + c\mbox{ for any }t\in[0,1]
$$  where $c\in\bbR$ is a constant and  $\xi$ satisfies the following:
    \be\label{e:sde_xi}
    d\xi_t = w(t,\xi_{t-}) dY_t^c + dC_t + J_t, \quad \xi_0 = 0 \quad a.s.,
    \ee
where
$$dC_t = \frac{w_x(t,\xi_{t-})}{2}w(t,\xi_{t-}) \left(d[Y,Y]_t^c-dt\right),$$
$$J_t = K_w^{-1}(t,K_w(t,\xi_{t-})+\Delta Y_t)-\xi_{t-}, \quad K_w(t,x)=\int_0^x\frac{1}{w(t,y)}dy+\int_0^t\frac{w_x(s,0)}{2}ds.$$

This pricing rule was initially proposed in Cetin, Danilova (2021) as a generalization of previous pricing rules, which does not lead to infinite insider profit. Moreover, we can notice that if insider strategy is absolutely continuous, then this SDE will have the form
$$d\xi_t = w(t, \xi_t)dY_t.$$

In above  $w$ is called weighting function which satisfies admissibility conditions of Definition (\ref{hwadm}).  
These admissibility conditions, together with the ones imposed on $\theta$ in Definition (\ref{thetaadm}) will ensure that SDE (\ref{e:sde_xi}) admits a unique strong Markov solution.

 Consider $P^{0,z}$ -- the time $0$ law of the process $(\xi,Z)$ starting from $(0,z)$. Then the market maker's measure $\bbP$, defined on $\left(\Omega, \cF_1^{Y,Z}\right)$, is given by 
\begin{equation*}\label{def:P}
\bbP(E) = \int_\bbR P^{0,z}(E) \mu(dz), \quad \forall E \in \cF_1^{Y,Z}. 
\end{equation*}

\textbf{\emph{Insider}} observes the price process $P$ and signal process $Z$ up to any time $t$,  thus, her filtration is given by $\cF_t^I = \cF_1^{P,Z}$. Insider's objective is to maximise the expected utility of final wealth, i.e.:

\be\label{ins_max}
\sup_{\theta \in \cA(w,c)} \bbE^{0,z} \left[-\frac{1}{\gamma} \exp\left\{- \gamma W_1^\theta \right\}\right],
\ee

where $\gamma$ is given in (\ref{e:PDEa}) and $\cA(w,c)$ is the set of admissible trading strategies for pricing rule $(w,c)$ specified in Definition (\ref{thetaadm}). The expectation is taken under the measure $P^{0,z}$ defined above. 

We denote by $W^{\theta}_1$ an insider's wealth at terminal time if she chooses to follow the trading strategy $\theta$. It is comprised of the continuous gain over the time interval $[0,1)$ and gain from the possible price discrepancy at terminal time $t = 1$, i.e.
\begin{equation*}\label{wealth}
W_1^{\theta} =  \int_0^{1-} \theta_{t-} dP_t + (Z_1-P_{1-})\theta_{1-}.
\end{equation*}

\section{Admissibility and Equilibrium}
\label{Admissibility and Equilibrium}

The above market model suggests a feedback mechanism for the insider, as her trading strategy will be reflected upon the asset price which in turn will influence her trading strategy itself. In this paper, we focus on finding the equilibrium of such market model in the following sense: 
\begin{enumerate}
\item given the pricing rule, insider's trading strategy is optimal;
\item given the trading strategy, there exists a unique strong solution for SDE (\ref{e:sde_xi}) over $[0,1)$ and the pricing rule is rational, i.e., martingale over $[0,1)$.
\end{enumerate}
To formalize the definition of equilibrium and rational pricing, we need to define the sets of admissible pricing rules and admissible trading strategies.

\begin{definition}\label{hwadm} An admissible pricing rule is a measurable weighting function $w$ and a constant c:
\begin{enumerate}
\item $w$ is defined on $[0, 1]\times I$, where $I$ is given in Assumption \ref{a:Z}.
\item $w \in \cC^{1,2} \left( [0,1] \times I \right)$ and is positive.
\item The weighting function satisfies:
\be\label{e:pde_w}
\frac{w_t}{w^2}(t,\xi)+\frac{w_{\xi\xi}(t,\xi)}{2}=-\gamma.
\ee
\item There exists a unique strong solution $\xi$ to the SDE 
\begin{equation*}\label{e:sde_xi_ni}
    d\xi_t = w(t,\xi_t) dB_t, \quad \xi_0 = 0 \quad a.s.
\end{equation*}
 in $\left(\Omega, \cF, (\cF_t)_{t\in [0,1)}, \bbQ^{base}\right)$.
\end{enumerate}
\end{definition}

\begin{remark} It can be shown, following the methodology developed in \cite{CD-GKB}, that in order for the equilibrium to exist the weighting function $w$ should satisfy (\ref{e:pde_w}). Thus, the condition (3) is necessary for the existence of equilibrium.
\end{remark}
\begin{remark} The condition (4), in essence, states that the market maker should chose the weighting function such that the market price is well defined if insider refrains from trading.
\end{remark}

\begin{definition}\label{def:rational pricing} We will call an admissible pricing rule rational if it satisfies
\begin{equation*}
P_t = \bbE \left[ Z_t \left| \cF_t^Y \right.\right]   
\end{equation*}
for a given admissible trading strategy $\theta$.
\end{definition}

Next, we turn to the definition of insider's admissible strategy. The minimal requirement for the admissibility is that the market price is well defined, i.e. SDE (\ref{e:sde_xi}) has the unique strong solution. Whereas the definition of admissible pricing rule ensures that the market price is well defined in the absence of the insider trading, the insider can only choose a trading strategy that results in the unique price.  Thus, the set of insider's admissible strategies is determined by the admissible pricing rule chosen by the market maker. The formal definition is as follows.

 \begin{definition}\label{thetaadm}
Given an admissible pricing rule $w$, an admissible  insider's trading strategy (denoted as $\theta  \in \cA(w)$)  is $\cF^{\xi, Z}$  adapted process satisfying:
\begin{enumerate}
\item $\theta$ is a semi-martingale with summable jumps on the filtration produced by $B$ and $Z$.
\item
There exists a unique strong solution of SDE (\ref{e:sde_xi}) in $\left(\Omega, \cF, (\cF_t)_{t\in [0,1)}, \bbQ^{base}\right)$.
\item
$(\xi, Z)$ is a Markov process adapted to $(\cF_t)_{t\in [0,1)}$ with measure $P^{0,z}$;
\item 
$\bbE^{0,z}\left[\exp \left\{-\gamma\int_0^1 P_t dB_t-\frac{\gamma^2}{2}\int_0^1 P_t^2 dt\right\}\right] =1$.
\end{enumerate}
\end{definition}

\begin{definition}\label{equil}
A pair $((w, c), \theta)$ is an equilibrium if $(w, c)$ is an admissible pricing rule, $\theta$ is an admissible insider strategy and:
\begin{enumerate}
    \item Given $\theta$, $(w, c)$ is rational pricing rule (according to the Definition (\ref{def:rational pricing}));

    \item Given $(w, c)$, $\theta$ maximizes the expected utility of insider final wealth (\ref{ins_max}).
\end{enumerate}
\end{definition}

\section{General Result}
\label{Result0}
\begin{theorem} \label{th:main}
The equilibrium is given by $c=0$,
$$w(t,x)=a(t,x)$$
and 
 \begin{gather*}
        d\theta_t = \alpha_tdt, \hspace{5mm} \alpha_t = w(t, \xi_t) \frac{\rho_x(t, \xi_t, V(t), Z_t)}{\rho(t, \xi_t, V(t), Z_t)},
    \end{gather*}
    where $\rho$ is the transition density of the process $\eta$, solving (\ref{e:SDE_eta})
 \end{theorem}

We will prove the Theorem via a sequence of Lemmata that will establish the result. We start with the proof that the candidate equilibrium insider's strategy yields the unique strong solution for the  SDE governing market price process. Moreover, we will demonstrate that this price is fully revealing at time $1$. 
\begin{lemma}\label{l:SDExi} Let $\rho$ be the transition density of the process given by (\ref{e:SDE_eta}). 
The SDE
\be\label{e:SDExi}
 d\xi_t= w(t, \xi_t) \frac{\rho_x(t, \xi_t, V(t), Z_t)}{\rho(t, \xi_t, V(t), Z_t)}dt+w(t, \xi_t)dB_t,
\ee
 admits the unique strong solution in $\left(\Omega, \cF, (\cF_t)_{t\in [0,1)}, \bbQ^{base}\right)$ on $[0,1)$. Moreover, the solution  satisfies  $\xi_1=Z_1$ $\bbQ^{base}$-a.s.. Furthermore, 
 $$
    Y_t= \frac{\rho_x(t, \xi_t, V(t), Z_t)}{\rho(t, \xi_t, V(t), Z_t)}dt+dB_t
 $$
 is a Brownian Motion in the filtration $\left(\cF^\xi_t\right)_{t\in[0,1]}$.
\end{lemma}

\begin{proof}

Consider the following function 
    $$
      v(t,x)=\int_0^x\frac{1}{a(t,y)}dy+\int_0^t\frac{a_x(s,0)}{2}ds, \hspace{5mm} x \in I.
    $$
    Due to  the Assumption \ref{a:Z} the range of function $v$ is  $\bbR$. It is continuous and strictly increasing in $x$, thus, admits a continuous increasing inverse. Denote this inverse as  $\lambda: [0,1]\times\bbR\rar I$, i.e.
$$
   v(t,\lambda(t,y))=y\mbox{ and }\lambda(t,v(t,z))=z \mbox{ for any } y\in \bbR, z\in I.
$$

Let $\kappa_t = v(t, \eta_t)$ -- this process does not explode on $[0,1]$.
Indeed, suppose $\kappa$ explodes and consider the sequences of stopping times $\tau^u_n= \min \{t\geq 0: \kappa_{t} >n\}$ and $\tau^l_n= \min \{t\geq 0: \kappa_{t} <-n\}$. Let $\tau^i=\lim_{n\to\infty} \tau_n^i$, $i=l,u$. Then either $\bbQ^{base}[\tau^l\leq 1]>0$ or  $\bbQ^{base}[\tau^u\leq 1]>0$. Suppose, wlog, $\bbQ^{base}[\tau^u\leq 1]>0$, then for $\omega\in \{\tau^u\leq 1\}$ we will have
$$
  \eta_{\tau^u}= \lim_{n\to\infty}\eta_{\tau^u_n} = \lim_{n\to\infty}\lambda(\tau^u_n, \kappa_{\tau^u_n}) = \lim_{n\to\infty}\lambda(\tau^u_n, n) = \lambda(\tau^u, +\infty)=u,
$$ 
which contradicts the assumption that the domain of $\eta$ is $(l,u)$. Thus, we conclude that $\tau^u > 1$ a.s.. Similar arguments yield $\tau^l > 1$ a.s. and therefore $\kappa$ does not explode on $[0, 1]$. 

Direct application of Ito lemma yields 
\bean
    d\kappa_t &=& (v_t(t, \eta_t) +  \frac{a^2(t, \eta_t)}{2} v_{xx}(t, \eta_t))dt+  a(t, \eta_t) v_x(t, \eta_t)d\beta_t \\
     &=&d\beta_t + \left(\int_0^{\eta_t} \left(\gamma + \frac{a_{xx}(t, y)}{2}\right)dy + \frac{a_x(t,0)}{2} - \frac{a_x(t, \eta_t)}{2}\right)dt \\
     &=&d\beta_t + \gamma \eta_t dt.
\eean
Thus, $\kappa$ solves SDE
\begin{gather} \label{e:kappa_SDE}
    d\kappa_t = d\beta_t + \gamma\lambda(t, \kappa_t)dt.
\end{gather}
This SDE has the unique strong solution. The existence of strong solution is obvious as  $\kappa_t = v(t, \eta_t)$ is a solution. As for uniqueness, suppose  there is another strong solution $\tilde{\kappa_t}$. Consider a process $\tilde{\eta_t} = \lambda(t, \tilde{\kappa_t})$. Due to direct application of Ito lemma it will satisfy (\ref{e:SDE_eta}), so due to the Assumption \ref{a:Z} $\tilde{\eta}\equiv\eta$ and therefore $\tilde{\kappa}$ will coincide with $\kappa$ as $v$ is strictly monotone.

Next we will show that the SDE (\ref{e:kappa_SDE}) can be viewed as a weak conditioning of a Brownian Motion  (see, e.g. Theorem 4.1 from \cite{DMB-CD}) which will allow us to characterize its transitional density.

First, we define the function 
$$u(t, x) = \exp\left\{\int_0^x \gamma \lambda(t,y)dy - \int_0^t\left(\frac{\gamma\lambda_x(s, 0)}{2} +\frac{\gamma^2\lambda^2(s, 0)}{2}\right)ds + C_0\right\},
$$
where $C_0$ is some constant. 

This function  satisfies $\gamma\lambda(t, x) = \frac{u_x(t, x)}{u(t, x)}$, thus, we can rewrite
\begin{equation*}\label{eq:kappa_u}
d\kappa_t = d\beta_t +  \frac{u_x(t, \kappa_t)}{u(t, \kappa_t)}dt.
\end{equation*}
It is evident that $u(t, x) \in \cC^{1,2} ([0,1), \bbR)$ and is strictly positive in this domain.

Moreover, $u(t, \tilde{B}_t)$, where $\tilde{B}$ is a Brownian motion, is a true martingale. Indeed, due to the Lemma \ref{l:mart} the fact there exists unique strong solution to (\ref{e:kappa_SDE}) and $d\tilde{\kappa}_t=d\beta_t$ implies that 
$$
  L_t=\exp\left\{\int_0^t\frac{u_x(s,\beta_s)}{u(s,\beta_s)}d\beta_s-\frac{1}{2}\int_0^t\frac{u^2_x(s,\beta_s)}{u^2(s,\beta_s)}ds\right\}
$$
is a martingale. 

Direct calculations yield
$$
\lambda_t(t,x) + \frac{1}{2}\lambda_{xx}(t,x) = -\gamma\lambda(t,x) \lambda_x(t,x)
$$
and
\begin{gather*}
    u_t(t, x) + \frac{u_{xx}(t, x)}{2} = 0,
\end{gather*}
thus,
$$
d\log(u(s,\beta_s))=\frac{u_x(s,\beta_s)}{u(s,\beta_s)}d\beta_s-\frac{1}{2}\frac{u^2_x(s,\beta_s)}{u^2(s,\beta_s)}ds
$$
implying 
$$
  L_t=\frac{u(t,\beta_t)}{u(0,0)}
$$
which leads to the claimed result.

Similarly, an application of Lemma \ref{l:mart} with the inverted roles of processes $\kappa$ and $\tilde{\kappa}$ yield that $\frac{1}{u(t,\kappa_t)}$ is a true martingale.

Since $u(t, x)$ satisfies the conditions of  $h$-function in the Theorem 4.1 of \cite{DMB-CD}, the (\ref{e:kappa_SDE}) is indeed the SDE of a Brownian motion weakly conditioned by $u$. In particular, the transition density of the process $\kappa$, $p(.)$, satisfies
$$p(s, x; t, y)= \frac{1}{u(s, x)}u(t, y)\Gamma(s, x, t, y), $$
where $\Gamma$ is the transition density of a standard Brownian motion.

Consider the process
$$
  U_t=v(V(t),  Z_t)=v(V(t),\eta_{V(t)})=\kappa_{V(t)}.
$$
Direct change of time yields that SDE for $U$ is:
\begin{gather*} \label{U_SDE}
    dU_t = \sigma(t)d\tilde{\beta}_t + \gamma\lambda(V(t), U_t)\sigma^2(t)dt.
\end{gather*}
 where $\tilde\beta$ is defined in the Remark \ref{TimeCh}.

Now we are in the position to prove that the process 
\begin{gather} \label{R_SDE}
    dR_t = dB_t +  \gamma\lambda(t, R_t)dt + \frac{p_x(t, R_t, V(t), U_t)}{p(t, R_t, V(t), U_t)}dt,
\end{gather}
admits a unique strong solution and $R_1 = U_1$ $\bbQ^{base}$-a.s..

First, using the relation between $p(.)$ and the transition density of Brownian motion we can represent
\bean
    \frac{p_x(t, R_t, V(t), U_t)}{p(t, R_t, V(t), U_t)} &=& \frac{\Gamma_x(t, R_t, V(t), U_t)}{\Gamma(t, R_t, V(t), U_t)} -  \frac{u_x(t, R_t)}{u(t, R_t)} \\
    &=& \frac{U_t - R_t}{V(t) - t} - \frac{u_x(t, R_t)}{u(t, R_t)}.
\eean
Thus, the equation (\ref{R_SDE}) becomes:
\begin{gather*} \label{R_SDE_simp}
    dR_t = dB_t +   \frac{U_t - R_t}{V(t) - t} dt.
\end{gather*}
And therefore we have the following system of equations:
\begin{equation*}
    \begin{cases}
        dU_t = \sigma(t)d\tilde{\beta}_t + \gamma\lambda(V(t), U_t)\sigma^2(t)dt,
        \\
        dR_t = dB_t  +  \frac{U_t - R_t}{V(t) - t}dt.
    \end{cases}
\end{equation*}

Consider a new probability measure defined by
$$
\left.\frac{d\tilde{\bbP}}{d\bbQ^{base}}\right|_{\mathcal{F}^{\tilde{\beta},B}_1 }= \frac{u(v_0,U_0)}{u(1, U_1)} = \frac{(v_0,\kappa_{v_0})}{u(1, \kappa_1)} 
$$
which is an equivalent to $\bbQ^{base}$ since $ \frac{1}{u(t, \kappa_t)}$ is a true martingale due to the above considerations. This change of measure yields:

\begin{equation*}
    \begin{cases}
        dU_t = \sigma(t)d\beta^{\tilde{\bbP}}_t,
        \\
        dR_t = dB^{\tilde{\bbP}}_t + \frac{U_t - R_t}{V(t)-t}dt.
    \end{cases}
\end{equation*}

Due to Theorem 5.2  in  \cite{DMB-CD} this  system has the  unique strong solution and $R_1 = U_1$ $\tilde{\bbP}$-a.s.. Since the two measures are equivalent, the equation (\ref{R_SDE}) has the unique strong solution and $R_1 = U_1$ $\bbQ^{base}$-a.s., as claimed.

Now we are in position to establish the existence and uniqueness of solution of (\ref{e:SDExi}) and the fact that $\xi_1=Z_1$. Consider $\tilde{\xi}_t=\lambda(t,R_t)$ and observe that 
\bean
    d\tilde{\xi}_t &=& d\lambda(t, R_t) = \left(\lambda_t(t, R_t) + \frac{1}{2}\lambda_{xx}(t, R_t)\right)dt + \lambda_x(t, R_t)dR_t \\
             &=& -\gamma\lambda(t,R_t)\lambda_x(t, R_t)dt +  \lambda_x(t, R_t)dB_t + \gamma \lambda(t, R_t)\lambda_x(t,R_t)dt \\
              & & +\lambda_x(t, R_t)\frac{p_x(t, R_t, V(t), U_t)}{p(t, R_t, V(t), U_t)}dt \\
              &=&  w(t, \tilde{\xi}_t)dB_t + w(t, \tilde{\xi}_t)\frac{p_x(t, R_t, V(t), U_t)}{p(t, R_t, V(t), U_t)}dt
\eean
since
$$\lambda_x(t, R_t) = \lambda_x(t, \lambda^{-1}(t, \tilde{\xi}_t)) = \frac{1}{v_x(t, \lambda(t, \lambda^{-1}(t, \tilde{\xi}_t))} = w(t, \tilde{\xi}_t).$$
As $Z_t = \lambda(V(t), U_t)$,  $\tilde{\xi}_1 = Z_1$ $\bbQ^{base}$-a.s.. Moreover, as $\cF_t^R = \cF_t^{\xi}$ for $t \in [0, 1)$,
\bean
    p(t, R_t, V(t), z)dz &=& \bbP \left[ U_t \in dz \left| \cF_t^R \right.\right] = \bbP \left[ Z_t \in d\lambda(V(t), z) \left| \cF_t^{\xi} \right.\right] = \\
                                 &=& \rho(t,\tilde{ \xi}_t, V(t), \lambda(V(t), z))d\lambda(V(t), z) \\
                                 &=& \rho(t, \tilde{\xi}_t, V(t), \lambda(V(t), z))w(V(t), \lambda(V(t), z))dz \\
                                 &=& \rho(t, \lambda(t, R_t), V(t), \lambda(V(t), z))w(V(t), \lambda(V(t), z))dz.
\eean
That is,
\begin{gather*}
    \rho(t, y, V(t), z) = \frac{p(t, \lambda^{-1}(t, y), V(t), \lambda^{-1}(V(t), z))}{w(V(t), z)}.
\end{gather*}
Thus, 
\begin{gather}\label{eq:SDE_t_xi}
    d\tilde{\xi}_t = w(t, \tilde{\xi}_t)dB_t + w^2(t, \tilde{\xi}_t)\frac{\rho_x(t, \tilde{\xi}_t, V(t),  Z_t)}{\rho(t, \tilde{\xi}_t, V(t), Z_t)}dt,
\end{gather}

Since the SDE for $R_t$ has the unique strong solution, $\tilde{\xi}_t=\lambda(t, R_t)$, and  $\lambda(t, x)$ is continuous and strictly increasing in $x$, the SDE (\ref{eq:SDE_t_xi}) has the unique strong solution. Hence there exists the unique strong solution of (\ref{e:SDExi}) and $\xi_1=Z_1$ $\bbQ^{base}$-a.s., as claimed.

Finally, observe that
\begin{gather*}
    \bbE\left[\left.w(t, \xi_t) \frac{\rho_x(t, \xi_t, V(t), f(Z_t))}{\rho(t, \xi_t, V(t), f(Z_t))} \right|\cF_t^{\xi}\right] = \bbE\left[\left.\frac{p_x(t, R_t, V(t), U_t)}{p(t, R_t, V(t), U_t)} \right|\cF_t^{R}\right]
\end{gather*}
and 
\begin{gather*}
    \bbE\left[\left.\frac{p_x(t, R_t, V(t), U_t)}{p(t, R_t, V(t), U_t)} \right|\cF_t^{R}\right] = \int_{\mathbb{R}} p_x(t, R_t, V(t), u)du = 0, 
\end{gather*}
thus, 
$$
   \bbE\left[\left.\frac{\rho_x(t, \xi_t, V(t), f(Z_t))}{\rho(t, \xi_t, V(t), f(Z_t))} \right|\cF_t^{\xi}\right] =\frac{1}{w(t, \xi_t)}\bbE\left[\left.w(t, \xi_t) \frac{\rho_x(t, \xi_t, V(t), f(Z_t))}{\rho(t, \xi_t, V(t), f(Z_t))} \right|\cF_t^{\xi}\right] = 0
$$
and $Y$ is a Brownian Motion as claimed.
\end{proof}

\begin{lemma}\label{th:opt}
    Let  $(w,c)$ be a pricing rule satisfying  the  Definition \ref{hwadm}. Suppose there exists an absolutely continuous insider's strategy $\hat{\theta}\in \cA(w,c)$ such that $Z_1 = P_1$ a.s.. Then for any $\theta\in \cA(w,c)$ we will have
    \begin{gather*}
            E^{0,z} \left[ U\left(W_1^{\hat{\theta}}\right) \right] \geq E^{0,z} \left[ U\left(W_1^\theta\right) \right],
    \end{gather*}
    i.e. this insider's strategy is optimal. 
\end{lemma}
\begin{proof} Consider the function $\Psi^a(t, x)$:
\begin{gather*} \label{Psi_form_nl}
     \Psi^a(t, x) = \int_{a-c}^x \frac{u - (a-c)}{w(t, u)}du + \frac{1}{2} \int_t^1 w(s, a-c)ds.
\end{gather*}

As  $w(t,x)$ is positive we have
\begin{equation*}\label{eq:Psi_term}
\Psi^a(1-, x) \geq 0\mbox{ and } \Psi^a(1-, x) = 0 \Leftrightarrow x+c=a.
\end{equation*}

Moreover, 
\begin{equation}\label{eq:Psi_PDE}
\Psi_t^a(t, x) + \frac{w^2(t, x)}{2} \Psi_{xx}^a(t, x) = \frac{\gamma}{2} (x - (a-c))^2.
\end{equation}
Indeed, direct calculations yield
\begin{equation*}
\begin{cases}
    \Psi_t^a = -\int_{a-c}^x \frac{(u - (a-c))w_t(t, u)}{w^2(t, u)}du - \frac{1}{2} w(t, a-c) \\
    \Psi_{x}^a = \frac{x - (a-c)}{w(t, x)} \\
    \Psi_{xx}^a = \frac{1}{w(t,x)} - \frac{(x-(a-c))w_x(t, x)}{w^2(t, x)}.
\end{cases}
\end{equation*}

Thus,
\begin{gather*}
\Psi_t^a(t, x) + \frac{w^2(t, x)}{2} \Psi_{xx}^a(t, x) =\\
=-\int_{a-c}^x \frac{(u - (a-c))w_t(t, u)}{w^2(t, u)}du - \frac{w(t, a-c)}{2} + \frac{w(t,x)}{2} - \frac{(x-(a-c))w_x(t, x)}{2}.
\end{gather*}

Using Definition $\ref{hwadm}$ and integration by parts yields
$$
\begin{array}{ll}
-\int_{(a-c)}^x \frac{(u - (a-c))w_t(t, u)}{w^2(t, u)}du &= \int_{(a-c)}^x (u - (a-c))\left(\gamma + \frac{w_{xx}(t,u)}{2}\right)du\\
&= \gamma \int_{(a-c)}^x (u - (a-c))du+(x-(a-c))\frac{w_x(t, x)}{2} - \int_{(a-c)}^x \frac{w_x(t,u)}{2}du
\end{array}
$$
which establishes (\ref{eq:Psi_PDE}).
 
Next, applying Theorem 32 in \cite{Pro} to $\Psi^{a}(t, \xi_{t} )$ as well as (\ref{eq:Psi_PDE}) yields
\begin{gather*}
    W_1^{\theta} = \Psi^{Z_1}(0, 0) - \Psi^{Z_1}(1-, \xi_{1-}) -  \frac{1}{2} \int_0^{1-} \omega(t, \xi_{t-}) d[\theta, \theta]_t^c + \\
    + \sum_{0<t<1} \{\Psi^{Z_1}(t, \xi_t) - \Psi^{Z_1}(t, \xi_{t-}) - (\xi_{t} + c - Z_1)\Delta \theta_t\} + \\
    + \int_0^{1-} (\xi_{t-} + c - Z_1)dB_t + \int_0^{1-} \frac{\gamma}{2} (\xi_{t-} +c - Z_1)^2 dt,
\end{gather*}
in view of  the representation for the insider final wealth as
\begin{gather*}
        W_1^{\theta} = \int_0^{1-} (Z_1 - \xi_{t-} - c) d\theta_t - \int_0^{1-} w(t, \xi_{t-}) \{ d[B, \theta]_t + [\theta, \theta]_t^c \}.
\end{gather*}

Thus, the insider maximization problem becomes
\begin{gather*}
    1 + \sup_{\theta \in \mathcal{A}(\omega, c)} E^{0,z} \left[ -\frac{1}{\gamma}e^{-\gamma W_1^{\theta}} \right] = 1 - \frac{1}{\gamma} \inf_{\theta \in \mathcal{A}(\omega, c)} E^{0,z} \left[ e^{-\gamma \left(\Psi^{Z_1}(0, 0) - \Psi^{Z_1}(1-, \xi_{1-}) - M_{1-} + \sum_{0<t<1} D_t + \zeta_{1-}\right)} \right].
\end{gather*}
where 
\begin{gather*}
M_{1-} = \frac{1}{2} \int_0^{1-} w(t, \xi_{t-}) d[\theta, \theta]_t^c \geq 0,\\
 D_t = \Psi^{Z_1}(t, \xi_t) - \Psi^{Z_1}(t, \xi_{t-}) - (\xi_t +c - Z_1)\Delta \theta_t,\\
 \zeta_t = \int_0^t (\xi_{s-} + c - Z_1)dB_s + \int_0^t \frac{\gamma}{2} (\xi_{s-} + c- Z_1)^2 ds.
\end{gather*}

Observe that
\begin{gather*}
    D_t = \int_{\xi_{t-}}^{\xi_t} \frac{u +c - Z_1}{w(t, u)}du - (\xi_t + c - Z_1)\Delta \theta_t \leq \\
    \leq(\xi_t + c - Z_1)\int_{\xi_{t-}}^{\xi_t} \frac{1}{w(t, u)}du - (\xi_t + c - Z_1)\Delta \theta_t = 0,
\end{gather*}
since at each jump at time $t-$ we have $\int_{\xi_{t-}}^{\xi_t} \frac{1}{w(t, u)}du = K_w(t,\xi_t)-K_w(t,\xi_{t-})=\Delta Y_t=\Delta \theta_t$ according to the chosen pricing rule.

Thus, in view of positivity of $M_{1-}$, we obtain 
\begin{equation}\label{eq:Ins_opt_est}
    1 + \sup_{\theta \in \mathcal{A}(\omega, c)} E^{0,z} \left[ -\frac{1}{\gamma}e^{-\gamma W_1^{\theta}} \right] \leq 1 - \frac{1}{\gamma} \inf_{\theta \in \mathcal{A}(\omega, c)} E^{0,z} \left[e^{-\gamma \Psi^{Z_1}(0, c)}  e^{-\gamma \zeta_{1-}} \right].
\end{equation}
Note that in the above the equality reached only if $\theta$ is absolutely continuous and $P_1=Z_1$ as this will imply $M_{1-}=0$ and $D_t\equiv 0$.

Consider a change of measure given by (recall that $P_t=\xi_t+c$)
$$
\frac{dQ^{0,z}}{dP^{0,z}}=e^{-\gamma \int_0^1P_tdB_t-\frac{\gamma^2}{2}\int_0^1P^2_tdt}
$$
It is an equivalent change of measure for any admissible strategy and under measure $Q^{0,z}$ the processes $d\hat{B}_t=dB_t+\gamma P_tdt$ and $\beta$ are independent Brownian motions. Thus,
\begin{gather*}
    1 - \frac{1}{\gamma} \inf_{\theta \in \mathcal{A}(\omega, c)} E^{0,z} \left[e^{-\gamma \Psi^{Z_1}(0, c)}  e^{-\gamma \zeta_{1-}} \right] = 1 - \frac{1}{\gamma} \inf_{\theta \in \mathcal{A}(\omega, c)} E^{Q^{0,z}} \left[e^{-\gamma \Psi^{Z_1}(0, c)+\gamma Z_1\hat{B}_1-\frac{\gamma^2}{2}Z_1^2}   \right]
\end{gather*}

Moreover,    
    \begin{multline*}
    1 - \frac{1}{\gamma} \inf_{\theta \in \mathcal{A}(\omega, c)} E^{Q^{0,z}} \left[e^{-\gamma \Psi^{Z_1}(0, c)+\gamma Z_1\hat{B}_1-\frac{\gamma^2}{2}Z_1^2}   \right] = \\
    = 1 - \frac{1}{\gamma} \inf_{\theta \in \mathcal{A}(\omega, c)} E^{Q^{0,z}} \left[e^{-\gamma \Psi^{Z_1}(0, c)}  E^{Q^{0,z}}\left[e^{\gamma Z_1\hat{B}_1-\frac{\gamma^2}{2}Z_1^2}\mid \mathcal{F}_0\vee\sigma(Z_1)\right] \right].      
    \end{multline*}
Observe that in the enlarged filtration $\hat{B}$ is a Brownian motion as it is independent of $\beta$. Hence $E^{Q^{0,z}}\left[\left.e^{\gamma Z_1\hat{B}_1-\frac{\gamma^2}{2}Z_1^2}\right| \mathcal{F}_0\vee\sigma(Z_1)\right]=1$ and distribution of $Z_1$ under $Q^{0,z}$ is the same as under $P^{0,z}$. Thus,  
$$
    1 - \frac{1}{\gamma} \inf_{\theta \in \mathcal{A}(\omega, c)} E^{0,z} \left[e^{-\gamma \Psi^{Z_1}(0, c)}  e^{-\gamma \zeta_{1-}} \right] =1 + E^{0,z} \left[ -\frac{1}{\gamma}e^{-\gamma \Psi^{Z_1}(0, c)} \right]
    $$  
    
    Combining this with (\ref{eq:Ins_opt_est}) yields
    $$
    1 + \sup_{\theta \in \mathcal{A}(\omega, c)} E^{0,z} \left[ -\frac{1}{\gamma}e^{-\gamma W_1^{\theta}} \right] \leq 1 + E^{0,z} \left[ -\frac{1}{\gamma}e^{-\gamma \Psi^{Z_1}(0, c)} \right]
    $$  
    
    Moreover, in view of discussion after equation (\ref{eq:Ins_opt_est}) we obtain 
    $$
     1 + E^{0,z} \left[ -\frac{1}{\gamma}e^{-\gamma W_1^{\hat{\theta}}} \right]=1 + E^{0,z} \left[ -\frac{1}{\gamma}e^{-\gamma \Psi^{Z_1}(0, c)} \right]
    $$
    for $\hat\theta$ in the statement of the Lemma. Comparing the last two equations completes the proof.
   \end{proof}
The above two lemmata establish the result of Theorem  \ref{th:main} provided that the strategy considered in Lemma \ref{l:SDExi} is admissible. Indeed, Lemma \ref{th:opt} proves that  an absolutely continuous admissible strategy such that $P_1=Z_1$ is optimal for the insider. On the other hand, Lemma \ref{l:SDExi} provides a constructive example of such a strategy. Thus, if this strategy is admissible, the proof of Theorem \ref{th:main} is complete. 

\begin{proof} of Theorem \ref{th:main}
Due to the above discussion to establish the result  it remains to show that the strategy $\alpha$ is admissible. Verification of admissibility will follow closely the proof of Lemma \ref{l:SDExi} and we will freely use the notation from it.
 
Observe that, due to the result of Lemma \ref{l:SDExi} the admissibility will follow once we demonstrate that
$$
1=\bbE^{0,z}\left[e^{-\gamma \int_0^1P_tdB_t-\frac{\gamma^2}{2}\int_0^1P^2_tdt}\right]
$$ 
where $P_t=\xi_t=\lambda(t,R_t)$ and $\xi$ solves (\ref{e:SDExi}).

Let us consider the process $X_t = -\gamma P_t = -\gamma \xi_t= -\gamma\lambda(t,R_t)$. We need to prove that Doleans-Dade exponential $\mathcal{E}(X)_t$ defines $Q^{0,z}$ --  probability measure equivalent to $P^{0,z}$.
We know that $P^{0,z}$ is already equivalent to the measure $\tilde{P}$ defined in Lemma \ref{l:SDExi} via  $\frac{u(0,0)}{u(1,U_1)}$, so it is sufficient to prove that the measure $Q^{0,z}$ is equivalent to the measure $\tilde{P}$.

Following the proof of the Lemma \ref{l:mart}, the required result is established once it is shown that the two dimensional SDEs
\begin{equation} \label{e:DynBBr}
    \begin{cases}
        dU_t = \sigma(t)d\beta^{\tilde{\bbP}}_t,\quad U_0=0,
        \\
        d\tilde{R}_t = dB^{\tilde{\bbP}}_t + \frac{U_t - \tilde{R}_t}{V(t)-t}dt, \quad R_0=0,
    \end{cases}
\end{equation}
and 
\begin{equation}\label{e:RU}
    \begin{cases}
        dU_t = \sigma(t)d\beta^{\tilde{\bbP}}_t, \quad U_0=0,
        \\
        dR_t = dB^{\tilde{\bbP}}_t + \frac{U_t - R_t}{V(t)-t}dt - \gamma\lambda(t, R_t)dt, \quad R_0=0.
    \end{cases}
\end{equation}
have unique strong solutions.

The first SDE has the unique strong solution on $[0,1]$ due to  Theorem 5.2  in  \cite{DMB-CD}. Moreover, this solution doesn't explode on the interval $[0,1]$.

 As to the second SDE, as  its coefficients are locally Lipschitz and locally bounded, it admits unique strong solution up to the explosion time in view of Theorem 2.8 in \cite{DMB-CD}. So, it is left to prove that there is no explosion on $[0,1]$.

To demonstrate that, let $\Omega_1\subseteq \Omega$ be a set of $\omega\in \Omega$ such that: 1) solution of (\ref{e:DynBBr}) exists and continuous, and 2) there exists continuous $R$ solving (\ref{e:RU}) until stopping time $\tau(\omega)$. Observe that 
$\tilde{\bbP}(\Omega_1)=1$. Fix any $\omega^*\in \Omega_1$ and suppose $\tau(\omega^*)<1$. We have that $u(t)=R_t-\tilde{R}_t$ is continuous and solves
$$
  u'(t) =  -\frac{u(t)}{V(t)-t}dt - \gamma\lambda(t, u(t)+\tilde{R}_t(\omega^*))dt, \quad u(0)=0
$$
on $[0,\tau(\omega^*))$ and therefore 
\be\label{e:uODE}
   (u^2(t))' =  -\frac{u^2(t)}{2(V(t)-t)}dt - \frac{\gamma}{2} u(t)\lambda(t, u(t)+\tilde{R}_t(\omega^*))dt 
\ee
on $[0,\tau(\omega^*))$.

Since $\lambda(t,x)$ is smooth enough, the function $n(t)$, defined by $\lambda(t,n(t)) = 0$, is continuous, and therefore bounded on [0,1].
Thus, $n(t)-\tilde{R}_t(\omega^*)$ is bounded on $[0,1]$ and therefore 
$$
   -\infty<N_*=\min_{t\in [0,1]}(n(t)-\tilde{R}_t(\omega^*))\leq \max_{t\in [0,1]}(n(t)-\tilde{R}_t(\omega^*))=N^*<\infty.
$$
As $\lambda(t,\cdot)$ is increasing, $x\lambda(t,x+\tilde{R}_t(\omega^*))>0$ on $x\in \bbR\backslash [-(N_*)^-,(N^*)^+]$, hence
$$
\min_{t\in[0,1], x\in \bbR}x\lambda(t,x+\tilde{R}_t(\omega^*))=\min_{t\in[0,1], x\in [-(N_*)^-,(N^*)^+]}x\lambda(t,x+\tilde{R}_t(\omega^*))=-C>-\infty
$$ 
and it follows from (\ref{e:uODE}) that on $[0,\tau(\omega^*))$
\bean\label{e:uODEneq}
   (u^2(t))' \leq  -\frac{u^2(t)}{2(V(t)-t)}dt +cdt
\eean
where $c=\frac{\gamma C}{2}$. Due to Gronwall's inequality we obtain
$$
 u^2(t) \leq  ce^{-\int_0^t\frac{1}{2(V(u)-u)}du}\int_0^t e^{\int_0^s\frac{1}{2(V(u)-u)}du}ds\leq \tilde{c}(\omega^*) \mbox{ for all } t\in [0,\tau(\omega^*)).
$$
Note that $\tilde{c}(\omega^*)$ is a finite constant as $V(t)-t$ is bounded away from zero on $t\in  [0,\tau(\omega^*)]$ as $\tau(\omega^*)<1$. As $\tilde{c}(\omega^*)$ does not depend on $t$ and $\tilde{R}_{\cdot}(\omega^*)$ in bounded on  $[0,1]$ it leads to contradiction. Therefore $\tau(\omega)\geq 1$ for all $\omega\in \Omega_1$.

 Thus, we are left to establish that the solution of (\ref{e:RU}) does not explode at $1$. 

To this end consider two processes
\begin{equation*}
    \begin{cases}
        dR^{n+}_t = dB^{\tilde{\bbP}}_t + \frac{U_t - R^{n+}_t}{V(t)-t}dt - \gamma\lambda(t, -n)dt.        \\
        dR^{n-}_t = dB^{\tilde{\bbP}}_t + \frac{U_t - R^{n-}_t}{V(t)-t}dt - \gamma\lambda(t,n)dt.
    \end{cases}
\end{equation*}
Observe that until $\tau_n=\tau_n^+\wedge \tau_n^-$, where $\tau_n^+=\inf\{t\geq 0: R^{n+}_t >n\}$ and $\tau_n^-=\inf\{t\geq 0: R^{n-}_t <n\}$ we have, in view of Theorem 2.9 in \cite{DMB-CD}, 
$R^{n-}_t \leq R_t \leq R^{n+}_t $ and therefore 
$$
R^{n-}_{t\wedge\tau_n} \leq R_{t\wedge\tau_n} \leq R^{n+}_{t\wedge\tau_n}
$$

Note that for

$$\tilde{R}^{n+}_t=R^{n+}_t + \gamma e^{-\int_0^t\frac{1}{V(s)-s}ds}\int_0^t\lambda(s, -n)e^{\int_0^s\frac{1}{V(u)-u}du}ds$$

and 

$$\tilde{R}^{n-}_t=R^{n-}_t + \gamma e^{-\int_0^t\frac{1}{V(s)-s}ds}\int_0^t\lambda(s, n)e^{\int_0^s\frac{1}{V(u)-u}du}ds$$

we obtain the following SDEs
\begin{equation*}
    \begin{cases}
        d\tilde{R}^{n+}_t = dB^{\tilde{\bbP}}_t + \frac{U_t - \tilde{R}^{n+}_t}{V(t)-t}dt.        \\
        d\tilde{R}^{n-}_t = dB^{\tilde{\bbP}}_t + \frac{U_t - \tilde{R}^{n-}_t}{V(t)-t}dt.
    \end{cases}
\end{equation*}

Due to the Theorem 5.2 in \cite{DMB-CD} we conclude that $\tilde{R}^{n+}_1 =\tilde{R}^{n-}_1=U_1$ and therefore $R^{n+}_1=R^{n-}_1=U_1$ (due to application of L'Hospital rule to  $e^{-\int_0^t\frac{1}{V(s)-s}ds}\int_0^t\lambda(t, .)e^{\int_0^s\frac{1}{V(u)-u}du}ds$) for any $n$. 

Thus, we have
$$
   1_{[\tau_n\geq 1]}U_1\leq 1_{[\tau_n\geq 1]}\lim_{t\rar 1}R_t\leq 1_{[\tau_n\geq 1]}U_1.
$$
As $\tilde{\bbP}[\lim_{n\rar \infty}\tau_n\geq1]=1$ we conclude that $\lim_{t\rar 1}R_t=U_1$ and in particular the solution of SDE does not explode on $[0,1]$.

\end{proof}

\section{Examples}

\subsection{Deterministic volatility of the signal}\label{ch:DVS}

Consider the signal of the form

$$dZ_t = \Sigma(t)d\beta_t,\quad Z_0 \sim  N(0,q),\quad \Sigma(t)\geq 0.$$

Suppose that $\Sigma(t)$ is continuously differentiable function on $[0,1]$, $q\in \mathbb{R}, \, q \geq 0$, which satisfies on $[0,1)$
$$q +\int_0^t\Sigma^2(s)ds  > \frac{t}{C(C+\gamma t)}$$
and
$$\Sigma(1) \neq \frac{1}{C} - \gamma q -\gamma\int_0^1\Sigma^2(s)ds$$
for
\be\label{e:C_det}
C = \frac{-\gamma+\sqrt{\gamma^2+\frac{4}{q+\int_0^1\Sigma^2(t)dt}}}{2} >0.
\ee
\begin{remark} 
    For instance, one of the signals satisfying the assumption stated above is the signal with constant volatility $\Sigma(t)=\Sigma$, for which
    $$\left(\Sigma^2t+q)(C\gamma t + C^2\right) > t, \quad t\in [0,1)$$.
    We can see that on the left side of this inequality we have parabola with roots $-\frac{q}{\Sigma^2}$ and $-\frac{C}{\gamma}$ and, moreover, we can notice that this inequality becomes equality at $t=1$.
    In this case the inequality is equivalent to the condition of the derivative of the left part being equal or less than 1 at $t=1$ or equivalently:
    $$2\Sigma^2 C \gamma + Cq\gamma + \Sigma^2 C^2 \leq 1.$$

    This in its turn, may be true for example if $\Sigma = q$ and is sufficiently small, which shows us that our initial assumption can be achieved.
\end{remark}

This signal can be rewritten in the form

$$dZ_t = \sigma(t)a(V(t))d\beta_t,\quad Z_0=\int_0^{V(0)}a(s)d\tilde{\beta}_s$$
where $\tilde{\beta}$ is independent of $\beta$, 

\begin{eqnarray*}
V(t) &=& \frac{1}{\frac{\gamma}{C} - \gamma^2 q -\gamma^2\int_0^t\Sigma^2(s)ds} - \frac{C}{\gamma}, \quad V(0) = \frac{qC^2}{1-\gamma qC} > 0 \\
\sigma(t) &=& \sqrt{V'(t)} = \frac{\Sigma(t)}{\frac{1}{C} - \gamma q -\gamma\int_0^t\Sigma^2(s)ds}\\
a(t) &=& \frac{1}{\gamma t+C}
\end{eqnarray*}
and $C$ is given by (\ref{e:C_det}). Those satisfy the assumption (\ref{a:Z}), since the above assumption on $\Sigma(t)$ gives us that $V(t)>t$ on $[0,1)$. Thus, it is left to check that applying L'Hopital's rule we get
$$\lim_{t \to 1} D^2(t)\Lambda(t)\log{\Lambda(t)} = 0.$$
\begin{eqnarray*}
\lim_{t \to 1} D^2(t)\Lambda(t)\log{\Lambda(t)} &=& \lim_{t \to 1} \frac{\log{\Lambda(t)}}{\frac{1}{D^2(t)\Lambda(t)}} = \\
\lim_{t \to 1} \frac{V(t)-t}{\frac{2}{1+\sigma^2(t)}-\frac{V(t)-t}{D^2(t)\Lambda(t)}} &=& 0,
\end{eqnarray*}
since again due to L'Hopital's rule in our assumptions we have
$$\lim_{t \to 1} \frac{V(t)-t}{D^2(t)\Lambda(t)} = \lim_{t \to 1} \frac{\frac{(\sigma^2(t)-1)D^2(t) +2D^2(t)}{D^4(t)}}{\frac{1+\sigma^2(t)}{D^2(t)}} = 1.$$

Our theorem states that in this case the equilibrium is given by $c=0$,
$$
 w(t,x)=a(t,x) = \frac{1}{\gamma t + C}
$$
and 
 \begin{gather*}
        d\theta_t = \alpha_tdt, \hspace{5mm} \alpha_t = w(t, \xi_t) \frac{\rho_x(t, \xi_t, V(t), Z_t)}{\rho(t, \xi_t, V(t), Z_t)},
\end{gather*}

$\rho()$ is transition density of the process given by
$$d\eta_t = \frac{1}{\gamma t + C}d\beta_t.$$

Denoting
$$G(s, t) = \int_s^t \frac{1}{(\gamma\tau + C)^2}d\tau = \frac{t - s}{(\gamma s + C)(\gamma t + C)}$$
we get
$$\rho(s, y, t, x) = \frac{1}{\sqrt{2 \pi G(s, t)}}e^{\frac{-(x-y)^2}{2G(s, t)}}.$$

\subsection{Quadratic volatility of signal} 
Consider the signal of the form
$$dZ_t = (-\delta Z_t^2+bZ_t+d)d\hat{\beta}_t, \quad 0<|\delta|<\gamma, \quad \frac{d}{\delta} > 0, \quad Z_0 = \eta_{t_0},$$

where $t_0 = 1-\frac{\delta^2}{\gamma^2}$ and $\eta$ is the unique strong solution of 
$$
d\eta_t = \left(-\gamma \eta_t^2+\frac{\gamma b}{\delta}\eta_t+\frac{\gamma d}{\delta}\right)d\beta_t, \quad \eta_0=0.
$$

We can notice that it is possible to represent this signal as
$$dZ_t = \sigma(t)a(V(t),Z_t)d\beta_t$$
where we set $\beta_t = sign(\delta)\hat{\beta}_t$ and denote
\begin{eqnarray*}
V(t) &=& \frac{\gamma^2 - \delta^2}{\gamma^2} + \frac{\delta^2}{\gamma^2}t \\
\sigma(t) &=& \frac{|\delta|}{\gamma}\\
a(t,x) &=& -\gamma x^2 + \frac{\gamma b}{\delta}x + \frac{\gamma d}{\delta}.
\end{eqnarray*}

Direct calculations show that this  will satisfy the Assumption \ref{a:Z}.
Thus, due to Theorem \ref{th:main} the equilibrium is given by $c=0$,
$$
 w(t,x)=a(t,x) = -\gamma x^2 + \frac{\gamma b}{\delta}x + \frac{\gamma d}{\delta}
$$
and 
 \begin{gather}
        d\theta_t = \alpha_tdt, \hspace{5mm} \alpha_t = w(t, \xi_t) \frac{\rho_x(t, \xi_t, V(t), Z_t)}{\rho(t, \xi_t, V(t), Z_t)},
    \end{gather}
where $\rho$ is transition density of the process $Z$. An explicit form of this density function can be seen as Expression (11) in Ingersoll, "Valuing Foreign Exchange Rate Derivatives with a Bounded Exchange Process".

\subsection{Static insider signal}

$$Z_t = Z_1 = \eta_1,$$

where $\eta$ is the unique strong solution of
\bean
    \eta_t = \int_0^t a(s, \eta_s)d\beta_s. 
\eean
and
$a(t,x)$ satisfies assumption (\ref{a:Z}).

\begin{remark}
    It can be noticed that the two different cases of such $Z_1$ have already been described in the literature:
    \begin{enumerate}
        \item The case of static bounded $Z_1$ was described in the work of Shi (2013),
        \item The case of static $Z_1$, where $\gamma$ for $a(t,x)$ is sufficiently small, has been described in the work of Bose, Ekren (2023).
    \end{enumerate}
\end{remark}
In this case we can consider as a new insider signal $\tilde{Z_t} = \eta_{V(t)}$ for some $V(t)$ satisfying Assumption 2.1 and base insider strategy on $\tilde{Z_t}$. Thus, obtained insider signal will satisfy the Theorem 4.1.

This shows us that there can be achieved  equilibria, each for different $V(t)$.
These equilibria according to the achieved results will only differ by the insider strategy, but will have same weighting function and same ultimate benefit for the insider.

\begin{remark}
Though the example of static insider signal does not formally satisfy our assumptions for the main theorem due to $\sigma(t) = 0$ for this case, it is possible to directly apply the same approach.

First, let us take $V(t)=1$ so it will satisfy other parts of the assumption (\ref{a:Z}), in which terms
\begin{eqnarray*}
D(t) &=& 1-t \\
\Lambda(t) &=& \frac{t}{1-t}\\
\lim_{t \to 1} D^2(t)\Lambda(t)\log{\Lambda(t)} &=& 0.
\end{eqnarray*}

Second, we can notice that a new probability measure defined in standard setting by
$$
\left.\frac{d\tilde{\bbP}}{d\bbQ^{base}}\right|_{\mathcal{F}^{\tilde{\beta},B}_1 }= \frac{u(v_0,U_0)}{u(1, U_1)} = \frac{(v_0,\kappa_{v_0})}{u(1, \kappa_1)} 
$$
will be unnecessary and the Theorem 5.2 from \cite{DMB-CD} can be applied directly. 

The final substantial difference from the standard apporoach in the case of static insider signal will consist in the application of Lemma 6.1 for the SDEs
\begin{equation*}
    \begin{cases}
        dU_t = \sigma(t)d\beta^{\tilde{\bbP}}_t,
        \\
        dR_t = dB^{\tilde{\bbP}}_t + \frac{U_t - R_t}{V(t)-t}dt,
    \end{cases}
\end{equation*}
and 
\begin{equation*}
    \begin{cases}
        dU_t = \sigma(t)d\beta^{\tilde{\bbP}}_t,
        \\
        dR_t = dB^{\tilde{\bbP}}_t + \frac{U_t - R_t}{V(t)-t}dt - \gamma\lambda(t, R_t)dt.
    \end{cases}
\end{equation*}
Now, since $\sigma(t)=0$, it will be enough to apply the Lemma 6.1 to
Following the proof of the Lemma \ref{l:mart}, the required result is established 
$$dR_t = dB_t + \frac{U_1 - R_t}{V(t)-t}dt,$$
and 
$$dR_t = dB_t + \frac{U_1 - R_t}{V(t)-t}dt - \gamma\lambda(t, R_t)dt.$$

\end{remark}

\section{Appendix}
\subsection{Auxiliary results for main theorem}
Here we present some auxiliary result that is required to prove the lemmas that establish the statement of the main theorem.

\begin{lemma}\label{l:mart}
    Consider filtered probability space $\left(\Omega, \mathcal{F}, \{\mathcal{F}_t\}_{t\in [0,1]}, \mathbb{Q}^{base}\right)$ rich enough to support a $d$-dimensional Brownian motion $B$. Suppose that the following two SDEs
    \begin{equation} \label{SDE1}
        dX_t = \sigma(t)dB_t + \mu^X(t,X_t)dt,
    \end{equation}
    \begin{equation} \label{SDE2}
        dY_t = \sigma(t)dB_t + \mu^Y(t,Y_t)dt
    \end{equation}
    have unique strong solution on $[0,1]$, where $\mu^{X,Y}$ are $d$-dimensional continuous column functions on  $[0,1)\times \bbR^d$ and $\sigma$ is continuous $d*d$-matrix on  $[0,1)\times \bbR^d \times \bbR^d$.  Assume further that
    $$
      \mu(t,x):=\mu^Y(t,x)-\mu^X(t,x)
    $$ 
    is a  $d$-dimensional continuous column function on  $[0,1]\times \bbR^d$ and $\alpha^{-1}(t)$ is continuous $d*d$-matrix on  $[0,1)\times \bbR^d \times \bbR^d$ for $\alpha_{ij}(t)=\sum_{k=1}^d\sigma_{ik}(t)\sigma_{kj}(t)$. Then
    $$
      L_t=\exp\left\{\int_0^t(\alpha^{-1}(s)\mu(s, X_s))^T\sigma(s)dB_s-\frac{1}{2}\int_0^t(\alpha^{-1}(s)\mu(s, X_s))^T\mu(s)ds\right\}
    $$
    is a martingale on $[0,1]$.
\end{lemma}

\begin{proof}
   Consider a canonical filtered space $(C([0,1],\bbR^d), (\cB_t)_{t\in[0,1]}, \cB_1)$ and 2 infinitesimal generators associated with the 2 systems of SDEs:

    $$A^X_{t} = \frac{1}{2}\sum_{i,j}^d\alpha_{ij}(t)\frac{\partial^2}{\partial x_i \partial x_j } + \sum_{i=1}^d\mu^{X}_{i}(t,x)\frac{\partial}{\partial x_i},$$

    $$A^Y_{t} = \frac{1}{2}\sum_{i,j}^d\alpha_{ij}(t)\frac{\partial^2}{\partial x_i \partial x_j } + \sum_{i=1}^d\mu^{Y}_{i}(t,x)\frac{\partial}{\partial x_i}.$$

   The martingale problems for $(A^X,\delta_0)$ and  $(A^Y,\delta_0)$ are well-posed, since both respective SDEs have a strong unique solution (see Corollary 2.5 in \cite{DMB-CD}). Denote the solutions of those martingale problems as $P^X$ and $P^Y$ respectively.

    In view of Theorem 3.3 in \cite{Ruf15}, due to continuity of $\mu$ and $\alpha^{-1}(t)$
    $$
      P^Y\left(\int_0^1(\alpha^{-1}(s)\mu(s, X_s))^T\mu(s)ds<\infty\right)=1
    $$
    the process
    $$
       \tilde{L}_t=\exp\left\{\int_0^t(\alpha^{-1}(s)\mu(s, X_s))^T(dX_s-\mu^X(s,X_s)ds)-\frac{1}{2}\int_0^t(\alpha^{-1}(s)\mu(s, X_s))^T\mu(s)ds\right\}
    $$
    is a martingale under $P^X$ on $[0,1]$.
    
    Next, consider the original filtered probability space $(\Omega, \mathcal{F}, \{\mathcal{F}_t\}_{t\in [0,1]}, \mathbb{Q}^{base})$. As $X$ is the strong solution of (\ref{SDE1}), $(X,B), (\Omega, \mathcal{F}, \mathbb{Q}^{base}),  \{\mathcal{F}_t\}_{t\in [0,1]}$ is also a weak solution of (\ref{SDE1}). Thus, due to the Corollary 2.3 in \cite{DMB-CD}, $P^X=\mathbb{Q}^{base}X^{-1}$. Therefore $L$ is indeed a martingale on $[0,1]$.
\end{proof}

\bibliographystyle{plainnat}
\bibliography{ref.bib}

@book {Pro,
	AUTHOR = {Protter, Philip E.},
	TITLE = {Stochastic integration and differential equations},
	SERIES = {Stochastic Modelling and Applied Probability},
	VOLUME = {21},
	NOTE = {Second edition. Version 2.1,
	Corrected third printing},
	PUBLISHER = {Springer-Verlag, Berlin},
	YEAR = {2005},
	PAGES = {xiv+419},
	ISBN = {3-540-00313-4},
	MRCLASS = {60-02 (60G44 60H05 60H10 60H20)},
	MRNUMBER = {2273672},
	MRREVIEWER = {Evelyn Buckwar},
	DOI = {10.1007/978-3-662-10061-5},
	URL = {http://dx.doi.org/10.1007/978-3-662-10061-5},
}

@book{DMB-CD,
	title={Dynamic Markov Bridges and Market Microstructure: Theory and Applications},
	author={{\c{C}}etin, Umut and Danilova, Albina},
	volume={90},
	year={2018},
	publisher={Springer}
}

@article{Kyle,
	title={Continuous auctions and insider trading},
	author={Kyle, Albert S},
	journal={Econometrica: Journal of the Econometric Society},
	pages={1315--1335},
	year={1985},
	publisher={JSTOR}
}

@article{Back92,
	title={Insider trading in continuous time},
	author={Back, Kerry},
	journal={The Review of Financial Studies},
	volume={5},
	number={3},
	pages={387--409},
	year={1992},
	publisher={Oxford University Press}
}

@article{BP98,
	title={Long-lived information and intraday patterns},
	author={Back, Kerry and Pedersen, Hal},
	journal={Journal of financial markets},
	volume={1},
	number={3-4},
	pages={385--402},
	year={1998},
	publisher={Elsevier}
}

@article{CCDdef,
	title={Equilibrium model with default and dynamic insider information},
	author={Campi, Luciano and {\c{C}}etin, Umut and Danilova, Albina},
	journal={Finance and Stochastics},
	volume={17},
	number={3},
	pages={565--585},
	year={2013},
	publisher={Springer}
}

@article{D,
	title={Stock market insider trading in continuous time with imperfect dynamic information},
	author={Danilova, Albina},
	journal={Stochastics An International Journal of Probability and Stochastics Processes},
	volume={82},
	number={1},
	pages={111--131},
	year={2010},
	publisher={Taylor \& Francis}
}

@article{BCWmult,
	title={Imperfect competition among informed traders},
	author={Back, Kerry and Cao, C Henry and Willard, Gregory A},
	journal={The journal of finance},
	volume={55},
	number={5},
	pages={2117--2155},
	year={2000},
	publisher={Wiley Online Library}
}

@article{CCDbp,
	title={Dynamic Markov bridges motivated by models of insider trading},
	author={Campi, Luciano and Cetin, Umut and Danilova, Albina},
	journal={Stochastic Processes and their Applications},
	volume={121},
	number={3},
	pages={534--567},
	year={2011},
	publisher={Elsevier}
}

@article{HSmult,
	title={Long-lived private information and imperfect competition},
	author={Holden, Craig W and Subrahmanyam, Avanidhar},
	journal={The Journal of Finance},
	volume={47},
	number={1},
	pages={247--270},
	year={1992},
	publisher={Wiley Online Library}
}

@article{CD-GKB,
author = {\c{C}etin, Umut and Danilova, Albina},
title = {On Pricing Rules and Optimal Strategies in General Kyle--Back Models},
journal = {SIAM Journal on Control and Optimization},
volume = {59},
number = {5},
pages = {3973-3998},
year = {2021},
doi = {10.1137/20M1319267},
URL = {https://doi.org/10.1137/20M1319267},
eprint = {https://doi.org/10.1137/20M1319267}
}

@article{choRA,
	title={Continuous auctions and insider trading: uniqueness and risk aversion},
	author={Cho, Kyung-Ha},
	journal={Finance and Stochastics},
	volume={7},
	number={1},
	pages={47--71},
	year={2003},
	publisher={Springer}
}

@article{BERA23,
author = {Shreya Bose and Ibrahim Ekren},
title = {{Kyle–Back models with risk aversion and non-Gaussian beliefs}},
volume = {33},
journal = {The Annals of Applied Probability},
number = {6A},
publisher = {Institute of Mathematical Statistics},
pages = {4238 -- 4271},
year = {2023},
doi = {10.1214/22-AAP1905},
URL = {https://doi.org/10.1214/22-AAP1905}
}

@article{BERA24,
    author = {Bose, Shreya and Ekren, Ibrahim},
    title = {Multidimensional Kyle–Back Model with a Risk Averse Informed Trader},
    journal = {SIAM Journal on Financial Mathematics},
    volume = {15},
    number = {1},
    pages = {93-120},
    year = {2024},
    doi = {10.1137/21M1457059},
    URL = {https://doi.org/10.1137/21M1457059},
    eprint = {https://doi.org/10.1137/21M1457059}
}

@article{Bar02,
	title={Insider trading and risk aversion},
	author={Baruch, Shmuel},
	journal={Journal of Financial Markets},
	volume={5},
	number={4},
	pages={451--464},
	year={2002},
	publisher={Elsevier}
}

@article{Ruf15,
author = {Johannes Ruf},
title = {{The martingale property in the context of stochastic differential equations}},
volume = {20},
journal = {Electronic Communications in Probability},
number = {none},
publisher = {Institute of Mathematical Statistics and Bernoulli Society},
pages = {1 -- 10},
year = {2015},
doi = {10.1214/ECP.v20-3449},
URL = {https://doi.org/10.1214/ECP.v20-3449}
}

\end{document}